%% file: Main.tex
\documentclass[journal]{IEEEtran}

\usepackage{cite}
\usepackage[tbtags]{amsmath}
\usepackage{hyperref}

\usepackage{amsthm}
\usepackage{amssymb}
\usepackage{algorithm}
\usepackage{algpseudocode}
\usepackage{graphics,graphicx,subfigure}
\usepackage{epsfig}
\usepackage{caption}
\usepackage{wasysym}
\usepackage{multirow}
\usepackage{bm,color}
\usepackage{accents}
\usepackage{mathrsfs}
\usepackage{cleveref}
\usepackage{nomencl}
\graphicspath{{Figures/}}

\usepackage{hyperref}
\hypersetup{
    pdftitle={Tool for Improving Resilience on Electric Distribution Systems with Networked Microgrids},
    pdfauthor={Arthur Barnes},
    pdfsubject={Electric Power Distribution},
    pdfkeywords={electric distribution,resilience,mixed-integer optimization},
    bookmarksnumbered=false,     
    bookmarksopen=true,         
    bookmarksopenlevel=2,       
    colorlinks=false,            
    pdfstartview=Fit,           
    pdfpagemode=UseOutlines,    
    pdfpagelayout=TwoPageRight
}
\usepackage{hypcap}

\newcommand{\ubar}[1]{\underaccent{\bar}{#1}}

\newcommand{\poleFailureProb}{p}

\newcommand{\DGPerformancePeriod}{N}
\newcommand{\DGDepreciationRate}{\eta}
\newcommand{\DGYearlyRevenue}{r}

\newcommand{\SubProblem}{{\cal{Q}}(s)}
\newcommand{\microgrids}{{\mathcal{G}}}

\newcommand{\Nodes}{{\cal{N}}}
\newcommand{\Edges}{{\cal{E}}}
\newcommand{\Disasters}{{\cal{S}}}
\newcommand{\Damages}{{\cal{D}}}

\newcommand{\nodeVolts}{V}

\newcommand{\loadDemand}{\bar{\pmb{S}}_{l,k}^d}

\newcommand{\lineUseVariable}{b}

\newcommand{\hardenVariable}{h}

\newcommand{\facilityVariable}{u}

\newcommand{\lineExistsVariable}{e}
\newcommand{\lineDirectionVariable}{e}
\newcommand{\switchUseVariable}{w}
\newcommand{\lineCycleVariable}{\bar{b}}

\newcommand{\lineHardenVariable}{h}

\newcommand{\generatorVariable}{\pmb{S}_{i,k}^{gs}}
\newcommand{\loadVariable}{\pmb{S}_{i,k}^{ds}}
\newcommand{\loadServeVariable}{y}
\newcommand{\flowVariable}{\pmb{S}}

\newcommand{\lineCost}{c}

\newcommand{\hardenCost}{\psi}
\newcommand{\facilityCost}{\alpha}
\newcommand{\microgridCost}{\zeta}

\newcommand{\capacity}{\bar{S}}
\newcommand{\phases}{{\cal{P}}}
\newcommand{\phaseVariation}{\beta}
\newcommand{\availableCapacity}{\bar{\pmb{S}}_{l,k}^g}
\newcommand{\cycles}{\cal{C}}
\newcommand{\criticalLoad}{\lambda}
\newcommand{\load}{\gamma}
\newcommand{\Load}{\mathcal{L}}
\newcommand{\CritLoad}{\mathcal{K}}

\immediate\write18{makeindex \jobname.nlo -s nomencl.ist -o \jobname.nls}
\makenomenclature
\allowdisplaybreaks

\begin{document}
\title{Tools for Improving Resilience of Electric Distribution Systems with Networked Microgrids}
\author{Arthur~Barnes$^*$\thanks{$^*$Information Systems and Modeling (A-1 division), Los Alamos National Laboratory, NM, USA. {\tt\small (rbent@lanl.gov)}},~Harsha~Nagarajan$^{\dag}$,~Emre~Yamangil$^{\dag}$\thanks{$^\dag$Center for Nonlinear Studies, Los Alamos National Laboratory, NM, USA.},~Russell~Bent$^{*}$~and~Scott~Backhaus$^{*}$}


\maketitle

\begin{abstract}

In the electrical grid, the distribution system is the most vulnerable to severe weather events. Well-placed and coordinated upgrades, such as the combination of microgrids, system hardening and additional line redundancy, can greatly reduce the number of electrical outages during extreme events. Indeed, it has been suggested that resilience is one of the primary benefits of networked microgrids. We formulate a resilient distribution grid design problem as a two-stage stochastic program and make use of decomposition-based heuristic algorithms to scale to problems of practical size.  We demonstrate the feasibility of a resilient distribution design tool on a model of an actual distribution network. We vary the study parameters, i.e., the capital cost of microgrid generation relative to system hardening and target system resilience metrics, and find regions in this parametric space corresponding to different distribution system architectures, such as individual microgrids, hardened networks, and a transition region that suggests the benefits of microgrids networked via hardened circuit segments. 
\end{abstract}


\input{Sections/Intro}
\input{Sections/Nomenclature}
\input{Sections/Algo}
\input{Sections/NR}
\input{Sections/Conclusions}



\bibliographystyle{IEEEtran}
\bibliography{networked_ug_report}

\end{document}

%% file: Sections/Intro.tex
\section{Introduction}

Hurricane Sandy demonstrated shortcomings in the resilience of the North American power grid. Resilience is defined as the ability to operate the grid reliably under disaster conditions. Such disasters include earthquakes, tornadoes, ice storms, tsunamis, and wildfires \cite{wang_research_2016}. The transmission network is generally quite resilient because of redundancy from meshed lines and generation reserves. In contrast, the distribution grid is much less resilient. It is estimated that 80\% of all outages originate at the distribution level \cite{gonen2016electric}. This resilience is lower mainly because the networks are typically operated in a radial configuration, the components are more susceptible to damage, and there is little controllable generation in such networks \cite{ma_resilience_2016}. 

Hurricane Sandy caused prolonged outages for consumer, commercial, industrial and public loads alike. In response, the US Department of Energy has identified electric distribution grid resilience under severe weather as a research area to pursue, with the objective of improving resilience through advanced technologies such as microgrids, and has recognized the need for design tools that include resilience metrics \cite{ton_more_2015}. Microgrids are collections of co-located generation and loads that are capable of appearing to the main electrical grid as a single controllable load or generator, and which can be disconnected to operate as standalone grids \cite{shahidehpour_microgrids_2016}. Multiple options are available to improve resilience in distribution systems, including improved vegetation management, physical reinforcement of overhead structures, and addition of new lines and switching to increase circuit redundancy. Other potential options include reconfiguration, spare component stockpiling, improved scheduling of line crew dispatch, and installation of microgrids at critical customer locations \cite{wang_research_2016,ma_resilience_2016, shahidehpour_microgrids_2016}.


Several studies have investigated using microgrids to supply critical loads during extended outages on distribution systems\cite{arghandeh_local_2014,panteli_grid:_2015,abbey_powering_2014,bahramirad_building_2015,gholami_front_2016}. However, the cost of installing microgrids at all critical loads is prohibitive. We propose that microgrids exploit existing or upgraded network infrastructure and use switching to connect microgrids to each other or to other critical loads via hardened overhead or underground lines \cite{chen_resilient_2016,gao_resilience-oriented_2016}. Such a design philosophy takes advantage of the economies of scale and load diversity that make the bulk power grid cost-effective and more resilient than existing distribution grid design paradigms\cite{che15hierarchical}.

This manuscript describes a resilient design software tool (RDT) to determine an optimal design of distribution system upgrades combined with system operations to enable meeting resilience performance targets during extreme events. In our previous work \cite{yamangil2015resilient}, the approach was applied to small test systems using very simplified representations of power flow physics which neglected voltages, losses and reactive flows. In the current manuscript, the design tool is extended to use the  linearized approximation of unbalanced three-phase power flow and is applied to a very large model of an actual distribution network. 

To address this problem, we use a stochastic programming approach\cite{yamangil2015resilient,gao_resilience-oriented_2016} that requires the upgraded distribution network to meet minimum resilience requirements over many samples of widespread network damage corresponding to an extreme weather event. 
The formulation and algorithms for expanding and hardening distribution networks for resilience are similar to approaches used for transmission networks\cite{nagarajan_optimal_2016}; however, distribution systems present unique challenges. The first is that distribution networks are typically operated in a radial fashion. In the worst case, enforcing this topological constraint requires enumerating all cycles (loops) in the network and eliminating them systematically---an NP-complete problem \cite{valiant1979complexity}. In the literature of distribution system expansion planning, researchers have proposed various methods based on flow formulations to enforce radial requirements on small-scale, single-phase networks \cite{jabr2013polyhedral}. However, it is known in the optimization literature that flow-based formulations are computationally inefficient when compared with cut-based formulations \cite{nagarajan2015maximizing}. Thus, in this paper, we extend the latter to enforce radial operations in three-phase, large-scale networks. Second, it is challenging to handle the non-convex power flows of unbalanced multi-phase distribution networks in conjunction with topology design options. For a given radial topology, a number of relaxations and linear approximation methods have been proposed in the literature to make the nonconvex power flow equations tractable \cite{baran1989optimal,dallanese_distributed_2013,gan_convex_2014,sankur_linearized_2016}. Here, for unbalanced radial networks, we employ the linear approximation presented in \cite{gan_convex_2014,sankur_linearized_2016} but extend the method to handle the line on/off constraints via disjunctions. Further, we verify the quality of the solutions obtained with approximations using OpenDSS \cite{dugan2011open}, a comprehensive distribution system simulation tool, and demonstrate that the solutions are physically valid.  

\noindent{\textbf {Literature Review}} Recently, \cite{yuan_modified_2017} addressed the problem of placing microgrids for optimal restoration by topology re-configuration using a Viterbi algorithm. Their approach is a heuristic-based method that is limited to balanced systems and was demonstrated on small-scale networks (up to 69 buses). In a closely related paper \cite{ding_new_2017}, a method was developed to design resilient distribution systems by multiple microgrid formations (up to 615 buses). Their work focused on topology re-configuration of grids with balanced ``Lin-Dist" power flow models and do not consider upgrade options considered in this paper. Further, to enforce topology requirements, they consider the single-commodity flow formulation, which can have weak lower bounds\cite{nagarajan2015maximizing}. 
Related approaches such as\cite{yuan_robust_2016,ma_resilience_2016} use a defender-attacker-defender (DAD) model that is suited to finding a small set of critical upgrades that enable resilience to targeted attacks.  In principle, the DAD approach can be used for widespread extreme weather event damage, but, in practice, the computational complexity of a DAD approach grows quickly with the number of allowable failed components and is often intractable for our network design problem.  
To summarize, though there has been a great deal of interest in resilient design of distribution systems, there has not been much focus on upgrading existing large-scale grids subject to natural disasters with  practical design options as the respective mathematical formulation of this problem crosses the frontiers of tractability. Hence, in this paper, we address this issue by applying various state-of-the-art optimization techniques.




This rest of this manuscript is structured as follows: Section II describes the design criteria used as inputs to the resilient design problem, the fragility model used to generate sets of line outages based on a disaster scenario and the optimization problem formulation for resilient design. Section III describes the computer implementation of the resilient design tool and the case study systems it is applied to. Section IV presents results in terms of solutions produced by the RDT tool on the case study systems and variation in results by varying the user input parameters (as the trade-off between favoring construction of microgrids vs. construction of new lines). Section V presents conclusions and directions for future work.


%% file: Sections/Nomenclature.tex
\nomenclature[a000]{$\textbf{Parameters}$}{}
\nomenclature[a010]{$\Nodes$}{set of nodes (buses).}
\nomenclature[a020]{$\Edges$}{set of edges (lines and transformers).} 
\nomenclature[a030]{$\phases_{i}$}{set of phases allowed to consume or inject power at bus $i$.}
\nomenclature[a040]{$\phases_{ij}$}{set of phases for the line ($i,j$).}
\nomenclature[a050]{$\Load$}{set of all loads.}
\nomenclature[a060]{$\CritLoad$}{set of all critical loads.}
\nomenclature[a070]{$\mathcal{G}$}{set of existing and candidate microgrids.}
\nomenclature[a080]{$\Disasters$}{set of disaster scenarios.}
\nomenclature[a090]{$\Damages_s$}{set of edges that are inoperable during $s\in S$.}
\nomenclature[a100]{$\cycles$}{set of sets of nodes that includes a cycle.}


\nomenclature[a130]{$\loadDemand$}{complex power demand in kVA for load on phase $k$ of load $l$.}
\nomenclature[a130]{$\loadDemand$}{complex power demand in kVA for load on phase $k$ of load $l$.}
\nomenclature[a140]{$\capacity_{ij}^k$}{line capacity in kVA between bus $i$ and bus $j$ on phase $k$.}
\nomenclature[a150]{$\criticalLoad$}{fraction of critical load power that must be served.}
\nomenclature[a160]{$\load$}{fraction of total load power that must be served.}
\nomenclature[a170]{$\availableCapacity$}{generation capacity on phase $k$ of microgrid $l$ in kVA.}
\nomenclature[a175]{$\phaseVariation_{ij}$}{parameter for controlling how much variation in flow between the phases is allowed.}
\nomenclature[b0]{$\pmb{S}_{ij,0}^{k,s,l}$}{complex power point used to inner approximate thermal constraint $l$ on phase $k$ of line ($i,j$) during disaster $s$.}

\nomenclature[a180a]{${R^{k_1k_2}_{ij}}$}{resistance in $\Omega$ between phases $k_1$ and $k_2$ of line ($i,j$).}
\nomenclature[a180b]{${X^{k_1k_2}_{ij}}$}{reactance in $\Omega$ between phases $k_1$ and $k_2$ of line ($i,j$).}
\nomenclature[b190]{$\nodeVolts_{i,k}^s$}{magnitude-squared of voltage on phase $k$ in kV at node $i$ during disaster $s$.}
\nomenclature[a198]{$V_{min}$}{minimum magnitude-squared voltage in kV.}
\nomenclature[a198]{$V_{max}$}{maximum magnitude-squared voltage in kV.}
\nomenclature[a220]{$M$}{valid constant for disabling voltage constraints on nonexistent or open line ($i,j$).}

\nomenclature[a230]{$\lineCost_{ij}$}{cost in \$ to build line ($i,j$); 0 if line already exists.}
\nomenclature[a250]{$\hardenCost_{ij}$}{cost in \$ to harden line ($i,j$).}
\nomenclature[a270]{$\facilityCost_{l}$}{cost in \$ to build microgrid $l$.}

\nomenclature[a290]{$\DGPerformancePeriod$}{performance period of a microgrid in years.}
\nomenclature[a300]{$\DGDepreciationRate$}{depreciation rate of a microgrid.}
\nomenclature[a310]{$\DGYearlyRevenue$}{yearly revenue from operating a microgrid per kVA of installed capacity.}

\nomenclature[b000]{$\textbf{Variables}$}{}
\nomenclature[b010]{$\loadVariable$}{load delivered at bus $i$ on phase $k$ during disaster $s$.}
\nomenclature[b020]{$\loadServeVariable_l^s$}{determines if the $l$th load is served or not during disaster $s$.}
\nomenclature[b040]{$\generatorVariable$}{complex power generated at bus $i$ on phase $k$ during disaster $s$.}

\nomenclature[b050]{$\flowVariable_{ij}^{k,s}$}{$P_{ij}^{k,s} + \mathbf{i} Q_{ij}^{k,s}$:   complex power flow on phase $k$ of line ($i,j$) during disaster $s$.}

\nomenclature[b060]{$\lineExistsVariable_{ij}^s$}{determines if line ($i,j$) is used during disaster $s$.}
\nomenclature[b070]{$\lineDirectionVariable^s_{ij,0}$}{determines if flow exists on line ($i,j$) from $j$ to $i$ during disaster $s$.}
\nomenclature[b080]{$\lineDirectionVariable^s_{ij,1}$}{determines if flow exists on line ($i,j$) from $i$ to $j$ during disaster $s$.}
\nomenclature[b090]{$\lineUseVariable_{ij}$}{determines if line ($i,j$) is built.}
\nomenclature[b110]{$\hardenVariable_{ij}$}{determines if line ($i,j$) is hardened.}
\nomenclature[b120]{$\facilityVariable_{l}$}{determines if generator $l$ is constructed.}

\nomenclature[b140]{$\lineHardenVariable_{ij}^s$}{determines if line ($i,j$) is hardened during disaster $s$.}
\nomenclature[b150]{$\lineCycleVariable_{ij}^s$}{determines if at least one edge ($i,j$) is used during disaster $s$.}


\section{Nomenclature}
\renewcommand{\nomname}{}
Note that complex quantities (variables and constants) are denoted in bold. Given any two complex numbers $\pmb{z}_1$ and $\pmb{z}_2$, $\pmb{z}_1 \geq \pmb{z}_2$ implies $\Re(\pmb{z}_1) \geq \Re(\pmb{z}_2)$ and $\Im(\pmb{z}_1) 
\geq \Im(\pmb{z}_2)$.

\printnomenclature

%% file: Sections/Algo.tex
\section{Formulation and algorithms}
The high-level objective of the resilient design problem is to select a set of network upgrades that will reduce the maximum number of outages during 
an extreme event. This paper considers
severe ice and wind storms. The 
objective
selects the least expensive combination of possible upgrade options for a feeder, consisting of (i) adding new lines, (ii) hardening lines and (iii) adding microgrids that allows the feeder to meet resilience requirements. The resilience requirements are defined over a set of disaster scenarios. Each disaster scenario 
describes the power system after the event has disabled lines or other equipment.
The scenarios are the result of a stochastic calculation, in this case sampling from a probability distribution of pole failure due to an ice and wind storm \cite{sa_reliability_2002}.

\subsection{Metrics}
The design criteria for resilient design are (i) constraints for load served and (ii) costs for the upgrades. The problem is formulated to upgrade an electrical distribution system to meet load constraints at a minimal cost. Load constraints consist of critical load served and noncritical load served. The load served criteria are defined in terms of the total power supplied to critical and noncritical loads in each damage scenario.
In a typical design scenario, a large majority (90--100\%) of critical load is selected to be met and 10--50\% of noncritical load is selected to be met. The noncritical load constraint encourages the optimizer to find a solution that will help improve resiliency for ordinary customers while ensuring resilient power for critical loads. 

Without loss of generality, new lines are constructed underground at a cost that is linear with respect to their length \cite{ltps13utilization,house13placement}. Because the threat model is an ice and wind storm rather than flooding, new lines are perfectly reliable. See \cite{yamangil2015resilient} for how to model situations when new lines are not perfectly reliable.
To ensure that the distribution system is operated radially, new lines always come with a switch.

This work assumes that hardening lines involves removal of the existing overhead line and replacement with an underground line. As with new lines, hardened lines are also perfectly reliable. It is also assumed that the cost of removal of the overhead line is small compared to that of the new underground line, so the cost is the same as adding a new underground line, with the exception that a switch is not added. It is assumed that the existing feeder has sufficient switches to always operate in a radial configuration.

Each microgrid consists of generation connected directly to the critical load it serves plus a transfer switch connecting the microgrid to the primary system. The cost of installing a microgrid consists of a fixed cost plus a variable cost that is linear in the power rating of the microgrid generation. The fixed cost includes balance-of-system costs, the generator fuel tank, the transfer switch cost, and the fixed portion of the generator cost 
\cite{kurtz14backup}.

\subsection{Fragility Model}
This study is based on an ice and wind storm \cite{sa2002reliability}. The failure mode is ice accumulation on conductors and communications cables increasing the cross-sectional area such that wind pressure on the cables is sufficient to exceed the breaking strength of the pole. The probability of pole failure is assumed to be uniform for each pole in the system. This uniform pole failure model is used to trivially derive a line failure model in which each overhead line span can fail with the same probability. Given a failure probability $\poleFailureProb_i$ for each pole, the line failure probability, given by $1 - (1- \poleFailureProb_i)^2$, is the likelihood of one of the poles supporting a line span failing. For each line span, a weighted coin is flipped to determine whether the line fails in a particular disaster scenario. This is repeated for each disaster scenario. 

\subsection{Optimization for Design}

\begin{figure}[ht]
\centering
\includegraphics[scale=0.58]{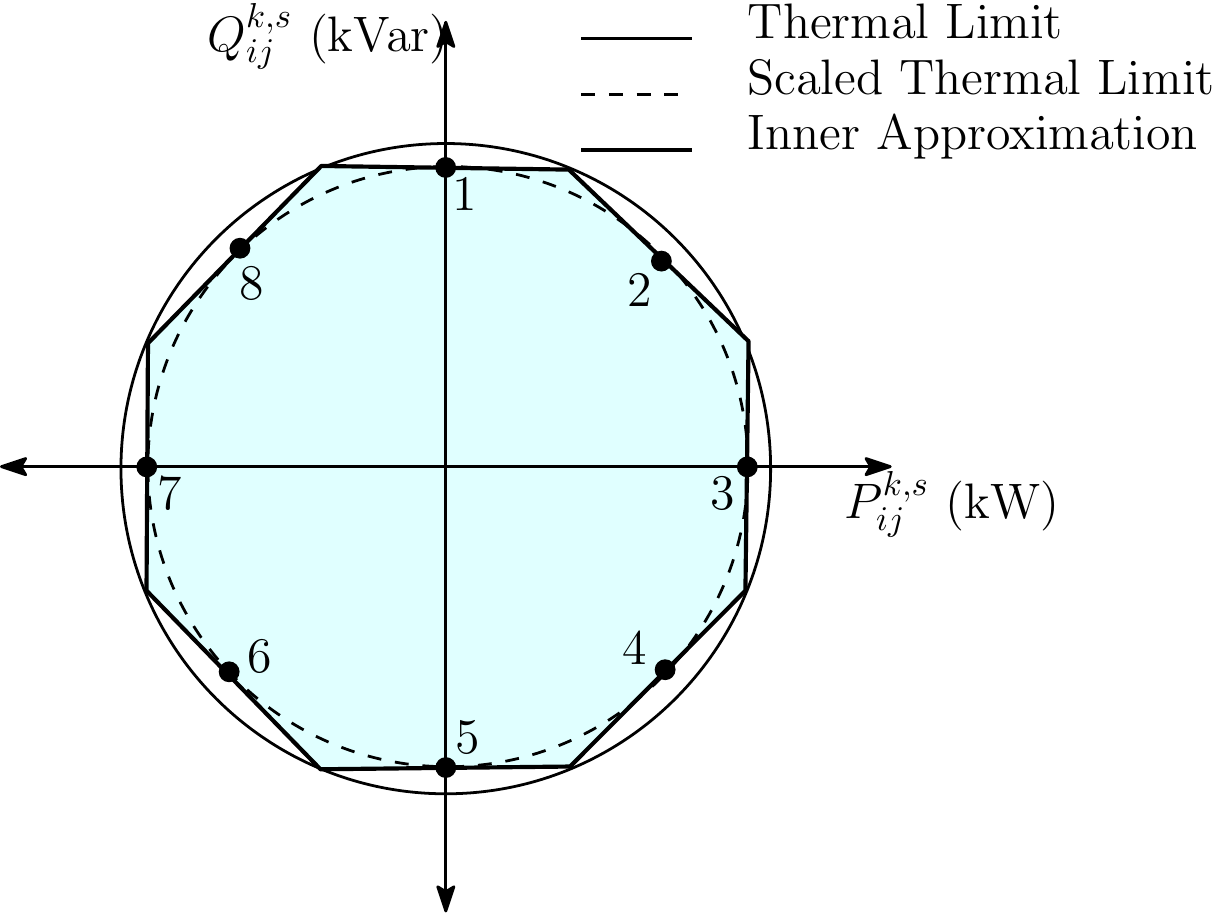}
\caption{Inner approximation of line thermal limit constraints.}
\label{fig:thermal_inner}
\end{figure}

The set of scenarios $S$ includes a set of disaster scenarios and the baseline scenario. Given a scenario $s\in S$, $\SubProblem$ in Eqs. \ref{eq:subprob_feas_set} and \ref{eq:subprob_voltage_cons} defines the set of feasible distribution networks.
In this model, Eqs. \ref{eq:line_thermal_cons_re} to \ref{eq:line_thermal_cons_diag} model an inner approximation of the capacity constraint on phase flows based on a polygon enclosed by eight lines. This approximation is illustrated in Figure \ref{fig:thermal_inner}. In the figure, the outer circle is the actual line thermal limit. The inner circle is the thermal limit scaled such that an octagon whose edges are tangent to the inner circle and whose vertices intersect the outer circle exists. This octagon describes a feasible region for $\SubProblem$, and the bold points mark the $(P_{ij,0}^{k,s,l},Q_{ij,0}^{k,s,l})$ points in \ref{eq:line_thermal_cons_re} to \ref{eq:line_thermal_cons_diag}.
When the line is not used, the flow is forced to 0 by $\lineExistsVariable_{ij}^s$.
Eq. \ref{eq:line_direction_cons} forces all phases to flow in the same direction, an engineering constraint.
Eq. \ref{eq:line_exists_cons} states that the flow on a line is 0 when the line is not available (switch open or not built).
Eq. \ref{eq:imbalance_cons} limits the fractional flow imbalance between the phases to a value smaller than $\phaseVariation_{ij}$.
Imbalance between phases cannot be extreme otherwise equipment may be damaged.
Here, we use $\phaseVariation_{ij}$ = 0.15 for transformers and $\phaseVariation_{ij}$ = 1.0 otherwise.
Eq. \ref{eq:damage_cons} states that the flow on a line is 0 when the line is damaged and not hardened (recall that only existing lines are damageable here).
Eq. \ref{eq:load_cons} requires all or none of the load at a bus to be served. Once again, this is an engineering limitation of most networks.
Eq. \ref{eq:gen_cons} limits the microgrid generation output by the generation capacity and caps the generation capacity that can be installed at each microgrid.
Eq. \ref{eq:balance_cons} ensures flow balance at the nodes for all phases. The summation  of $j \in \mathcal{N}$ collects edges oriented in the ``from'' direction connecting it’s neighboring nodes.
Eq. \ref{eq:cycle_cons} eliminates network cycles, forcing a tree or forest topology.
Eq. \ref{eq:crit_cons} ensures that a minimum fraction $\criticalLoad$ of critical load is served. 
Eq. \ref{eq:non_crit_cons} ensures that a minimum fraction of non critical load is served. 
Eqs.~\ref{eq:crit_cons} and \ref{eq:non_crit_cons} are the resilience criteria that must be met by  $\SubProblem$ and are similar to the $n-k-\epsilon$ criteria of \cite{yuan_optimal_2014}.
Eq. \ref{eq:discrete} states which variables are discrete.

\begin{subequations}
\fontsize{8}{8} \selectfont
\begin{equation} \begin{split} 
-\lineDirectionVariable_{ij,0}^s P_{ij,0}^{k,s,l} \leqslant P_{ij}^{k,s} \enskip \leqslant \lineDirectionVariable_{ij,1}^s P_{ij,0}^{k,s,l} \ \ \forall ij\in \Edges, k \in \phases_{ij}, l \in \{3, 7\}  \label{eq:line_thermal_cons_re} 
\end{split} \end{equation} 
\begin{equation} \begin{split} 
-\lineDirectionVariable_{ij,0}^s Q_{ij,0}^{k,s,l} \leqslant Q_{ij}^{k,s} \enskip \leqslant \lineDirectionVariable_{ij,1}^s Q_{ij,0}^{k,s,l} \ \ \forall ij\in \Edges, k \in \phases_{ij}, l \in \{1, 5\}  \label{eq:line_thermal_cons_im} 
\end{split} \end{equation} 
\begin{equation} \begin{split} 
P_{ij,0}^{k,s,l} P_{ij}^{k,s} + Q_{ij,0}^{k,s,l} Q_{ij}^{k,s} \enskip \leqslant ((P_{ij,0}^{k,s,l})^2 + (Q_{ij,0}^{k,s,l})^2) \\ \forall ij\in \Edges, k \in \phases_{ij}, l \in \{2, 4, 6, 8\}  \label{eq:line_thermal_cons_diag} 
\end{split} \end{equation} 
\begin{alignat}{2}
& \lineDirectionVariable_{ij,0}^s + \lineDirectionVariable_{ij,1}^s \leqslant \lineExistsVariable_{ij}^s & \quad\quad\quad\quad\quad\quad\quad\quad\quad\quad\quad\quad \forall ij\in \Edges & \label{eq:line_direction_cons} %
\end{alignat} 
\begin{equation} \begin{split}
\lineExistsVariable_{ij}^s = \lineUseVariable_{ij}^s & \quad \quad \quad \quad \quad  \quad \quad \quad \quad \quad\quad\quad\quad\quad\quad\forall ij\notin \Damages_s
\label{eq:line_exists_cons}
\end{split} \end{equation} 
\begin{alignat}{2}
& \frac{\displaystyle\sum_{k\in \phases_{ij}}\flowVariable_{ij}^{k,s}}{\frac{|\phases_{ij}|}{(1-\phaseVariation_{ij})}} \leqslant \pmb{S}_{ij}^{k^\prime,s} \leqslant \frac{\displaystyle\sum_{k\in \phases_{ij}}\flowVariable_{ij}^{k,s}}{\frac{|\phases_{ij}|}{(1+\phaseVariation_{ij})}}\quad & \forall ij\in \Edges, k^\prime \in \phases_{ij} & \label{eq:imbalance_cons}\\
& \lineExistsVariable_{ij}^s = \lineHardenVariable_{ij}^s  & \forall ij\in \Damages_s & \label{eq:damage_cons}\\
& \loadVariable = \sum_{l \in \mathcal{L}_i} \loadDemand \loadServeVariable_l^s & \forall i\in \Nodes, k\in \phases_i & \label{eq:load_cons}\\
& \generatorVariable \leqslant \sum_{l \in \mathcal{\microgrids}_i} \availableCapacity \facilityVariable_l & \forall i\in \Nodes, k\in \phases_i & \label{eq:gen_cons}\\
& \generatorVariable - \loadVariable - \sum_{j\in \Nodes} \flowVariable_{ij}^{k,s} = 0 & \forall i\in \Nodes, k\in \phases_i & \label{eq:balance_cons}\\
& \sum_{ij\in \Edges(C)} \lineExistsVariable_{ij}^s \leqslant |\Edges(C)|-1 & \forall C\in \cycles & \label{eq:cycle_cons}\\
%
& \sum_{i \in \Nodes(\CritLoad), k\in \phases_i} \loadVariable \geqslant \criticalLoad \sum_{l \in \CritLoad, k\in \phases_l} \loadDemand & & \label{eq:crit_cons}\\
& \sum_{i \in \Nodes, k\in \phases_i} \loadVariable \geqslant \load \sum_{l \in \Load, k\in \phases_l} \loadDemand & & \label{eq:non_crit_cons}\\
& \lineUseVariable_{ij}^s, \lineHardenVariable_{ij}^s, \switchUseVariable_{ij}^s, \lineExistsVariable_{ij}^s,\loadServeVariable_l^s, \facilityVariable_l \in \{0,1\} & & \label{eq:discrete}
\end{alignat}
\label{eq:subprob_feas_set}
\end{subequations}

Eqs. \ref{eq:line_z} to \ref{eq:voltage_phase_c_cons} define the linearized voltage drops across lines where the variables $V_{i,k}^{s}$ model the voltage magnitude squared on phase $k$ of node $i$ under scenario $s$ \cite{gan_convex_2014}. Eq. \ref{eq:voltage_range_cons} defines the allowable voltage range at each phase.

\begin{subequations}
\fontsize{8}{8} \selectfont
\begin{equation}
\begin{split}
\bar{\pmb{Z}}_{ij} &= Z_{ij}e^{\mathbf{i}2\pi/3}; \ubar{\pmb{Z}}_{ij} = Z_{ij}e^{-\mathbf{i}2\pi/3}, \mathrm{where,} \\
\bar{\pmb{Z}}_{ij}  &=    \bar{R}_{ij}+\mathbf{i}\bar{X}_{ij}, \ubar{\pmb{Z}}_{ij}  = \ubar{R}_{ij}+\mathbf{i}\ubar{X}_{ij}
\end{split}
\label{eq:line_z}
\end{equation}
%
\begin{equation}
M = V_{max} - V_{min} \quad \forall i\in \Nodes
\label{eq:big_m}
\end{equation}
%
\begin{equation} \begin{split}
-M(1 - \lineExistsVariable_{ij}^s) &\le V_{j,a}^{s} - V_{i,a}^{s} + 2(R_{ij}^{aa}P_{ij}^{a,s} + X_{ij}^{aa}Q_{ij}^{a,s} \\
&+ \bar{R}_{ij}^{ab}P_{ij}^{b,s} + \bar{X}_{ij}^{ab}P_{ij}^{b,s} + \ubar{R}_{ij}^{ac}P_{ij}^{c,s} + \ubar{X}_{ij}^{ac}Q_{ij}^{c,s}) \\  
&\le M(1 - \lineExistsVariable_{ij}^s) \quad \forall i\in \Nodes \label{eq:voltage_phase_a_cons} 
\end{split} \end{equation}
%
\begin{equation} \begin{split}
-M(1 - \lineExistsVariable_{ij}^s) &\le V_{j,b}^{s} - V_{i,b}^{s} + 2(\ubar{R}_{ij}^{ba}P_{ij}^{a,s} + \ubar{X}_{ij}^{ba}Q_{ij}^{a,s} \\
&+ R_{ij}^{bb}P_{ij}^{b,s} + X_{ij}^{bb}P_{ij}^{b,s} + \bar{R}_{ij}^{bc}P_{ij}^{c,s} + \bar{X}_{ij}^{bc}Q_{ij}^{c,s}) \\
&\le M(1 - \lineExistsVariable_{ij}^s) \quad \forall i\in \Nodes \label{eq:voltage_phase_b_cons} 
\end{split} \end{equation}
%
\begin{equation} \begin{split}
-M(1 - \lineExistsVariable_{ij}^s) &\le V_{j,c}^{s} - V_{i,c}^{s} + 2(\bar{R}_{ij}^{ca}P_{ij}^{a,s} + \bar{X}_{ij}^{ca}Q_{ij}^{a,s} \\
&+ \ubar{R}_{ij}^{cb}P_{ij}^{b,s} + \ubar{X}_{ij}^{cb}P_{ij}^{b,s} + R_{ij}^{cc}P_{ij}^{c,s} + X_{ij}^{cc}Q_{ij}^{c,s}) \\
&\le M(1 - \lineExistsVariable_{ij}^s) \quad \forall i\in \Nodes \label{eq:voltage_phase_c_cons} 
\end{split} \end{equation}
\begin{equation}
V_{min} \le V_{i,k}^{s} \le V_{max} \quad \forall i\in \Nodes, k\in \phases_i
\label{eq:voltage_range_cons}
\end{equation}
\label{eq:subprob_voltage_cons}
\end{subequations}

The cycle constraint of Eq. \ref{eq:cycle_cons} is implemented as Eqs. \ref{eq:no_cycle_1} to \ref{eq:line_implication_cons} to reduce the combinatorial growth of constraints.

\begin{subequations}
\fontsize{8}{8} \selectfont
\begin{alignat}{2}
& \sum_{ij\in \Edges(C)} (\lineCycleVariable_{ij}^s) \leqslant |\Edges(C)| - 1 & \quad\quad \forall C\in \cycles & \label{eq:no_cycle_1}\\
& \lineUseVariable_{ij}^s \leqslant \lineCycleVariable_{ij}^s & \forall ij \in \Edges & \label{eq:line_implication_cons} 
\end{alignat}
\label{eq:cycle_trick}
\end{subequations}

\noindent The number of cycles in the graph is significantly reduced by collecting each line between nodes into a single edge. A set of binary variables for the edges of the corresponding single-edge graph is introduced to enumerate the possible cycles in that graph (Eq. \ref{eq:no_cycle_1}).  Eq. \ref{eq:line_implication_cons} links the artificial cycle variables with the line and switch variables.

For each $s \in \Disasters$, $\SubProblem$ is the set of feasible networks. The optimal network upgrade is the minimum cost design that satisfies  $\SubProblem$  (Eqs. \ref{eq:obj} to \ref{eq:feasible_network}).

\begin{subequations}
\fontsize{8}{8} \selectfont
\begin{alignat}{2}
\min\quad &
\sum_{ij\in \Edges} ( \lineCost_{ij} \lineUseVariable_{ij} + \hardenCost_{ij} \hardenVariable_{ij})  
+ \sum_{l \in \mathcal{G}}  \facilityCost_{l} 
\facilityVariable_{l} \label{eq:obj}\\
\text{s.t.}\quad & & \nonumber\\
& \lineUseVariable_{ij}^s \leqslant \lineUseVariable_{ij} & \forall ij\in \Edges, s\in \Disasters & \label{eq:assign:1}\\
& \lineHardenVariable_{ij}^s = \hardenVariable_{ij} & \forall ij\in \Edges, s\in \Disasters & \label{eq:assign:3} \\
& \lineUseVariable_{ij}, \hardenVariable_{ij},  \facilityVariable_l \in \{0,1\} & \forall ij\in \Edges, l \in \mathcal{G} \label{eq:bin_vars} \\
& (\lineUseVariable_{ij}^s, \hardenVariable_{ij}^s,\facilityVariable_l) \in \SubProblem & \forall s\in \Disasters \label{eq:feasible_network}
\end{alignat}
\label{eq:optimization_model}
\end{subequations}

\noindent Eq. \ref{eq:obj} minimizes the cost of constructing new lines, hardening existing lines, and installing microgrids. The microgrids have fixed sizes and the cost of each microgrid is split into a fixed installation cost and a cost based on its capacity.
Eqs. \ref{eq:assign:1} through \ref{eq:feasible_network} link the first stage and second stage variables ($\SubProblem$). 
The inequality in Eqs. \ref{eq:assign:1} allows new lines to be switched.
Eq. \ref{eq:feasible_network} states that the vector of continuous and discrete variables $(\lineExistsVariable_{ij}^s,\loadServeVariable_l^s, \facilityVariable_l, \lineHardenVariable_{ij}^s)$ is a feasible network for scenario $s$. Note that $\lineUseVariable_{ij}^s=1$ for all existing lines in the network. Similarly, $\lineHardenVariable_{ij}^s=0$ for all candidate new lines.


The problem is solved using scenario-based decomposition (SBD) \cite{yamangil2015resilient}. SBD solves the mixed-integer network upgrade problem for the base case and a single scenario and then evaluates the designed network on the remaining scenarios to verify whether it is feasible. If the designed network is not feasible on any one of the remaining disaster scenarios, one of the infeasible scenarios is added to the upgrade problem and the process is repeated until feasibility is reached. SBD exploits the property of the problem that once the network is designed, the solutions for each scenario are independent of each other. 
A heuristic variable neighborhood search (VNS) is also applied to reduce computation time. VNS is used in the mixed-integer programs constructed by SBD.
VNS reduces the number of binary variables in the upgrade problem by fixing a subset of them based on their values resulting from a linear programming relaxation of the problem \cite{lazic_variable_2010}.



%% file: Sections/NR.tex
\section{Numerical Experiments}
The case studies in this paper are based on two feeders from a distribution system in the northeast United States that are relatively compact.
Each feeder is served by a different substation, and the feeders are meshed 
so that they can be backfed from substations (networked microgrids) on adjacent feeders.

\subsection{Implementation}

All results are based on a  C++ implementation using CPLEX 12.6.2 as the native mixed-integer linear programming solver. All studies were carried out on a server with 64 GB of memory and two 2.6 GHz Intel® Xeon® Processor E5-2660 v3 processors, where 32 of 40 available threads were used. The number of usable threads was limited by the CPLEX license.

\subsection{Case Study Systems}
\begin{figure}[ht]
\centering
\includegraphics[width=0.5\textwidth]{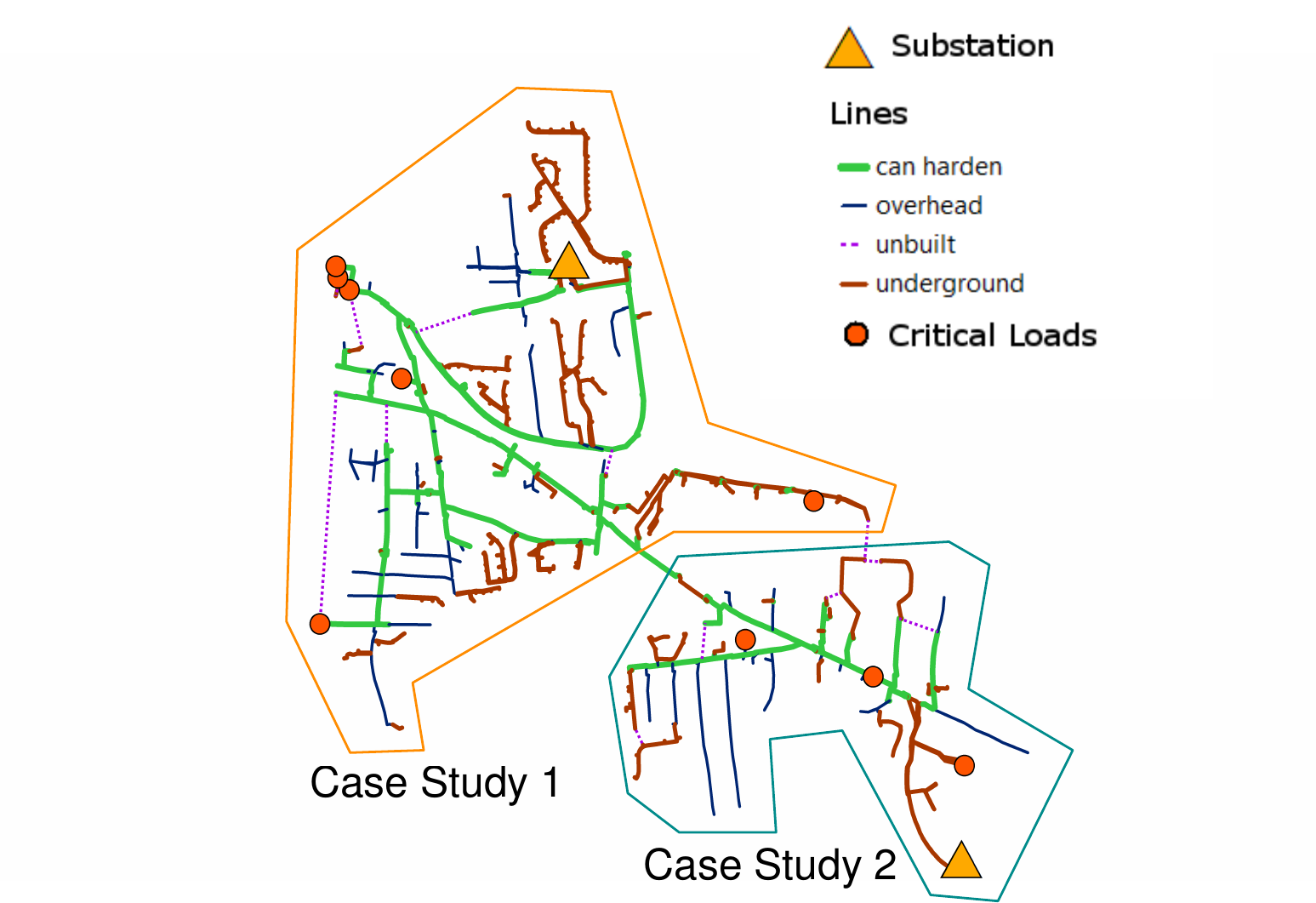}
\caption{Case Study 1 and Case Study 2 feeder.}
\label{fig:bn_setup}
\end{figure}

Three case studies are 
considered here.  The first two case studies consider each feeder independently. The third case study combines the two feeders.
The combined case study focuses on the advantages of meshing systems as networked microgrids.
Table \ref{tab:csprops} summarizes the salient characteristics of the two feeders, which are illustrated in Fig. \ref{fig:bn_setup}. The utility systems are unmodified with the exception that shunt capacitances are not included.

Case Study 1 focuses on the smaller of the two feeders. 
The feeder consists of two three-phase trunks connected to the substation by an underground line. It feeds a mix of mostly overhead and underground single-phase laterals with occasional three-phase laterals. The feeder in Case Study 1 serves a relatively small number of critical loads, including a mall that serves as an emergency shelter, a gas station near an interstate exit, and a small grocery store. 

Case Study 2 considers the larger of the two feeders. The critical loads this feeder serves consist of one cluster of three grocery stores and three pharmacies distributed throughout the feeder. The feeder in Case Study 2 has a long looping overhead three-phase trunk connected to a second main trunk following a major road. As in the feeder in Case Study 1, it serves a mix of mostly overhead and underground single-phase laterals. 

\begin{table}[t!]
\centering
\caption{Case Study System Properties} \label{tab:csprops}
\begin{tabular}{l r r}
\hline \hline
Parameter & Case Study 1 & Case Study 2  \\
\hline
Number of Buses & 271 & 945 \\
Number of Lines & 281 & 940 \\
Total Number of Loads & 125 & 333 \\
Total Load Power & 5.50 MW & 7.64 MW\\
Number of Critical Loads & 3 & 6 \\
Total Critical Load Power & 359 kW & 972 kW \\
Approximate Span East-West & 1.5 km & 2.1 km \\
Approximate Span North-South & 1.2 km & 2.3 km \\
\hline \hline
\end{tabular}
\end{table}


Table \ref{tab:psprops} summarizes the upgrade options for the three case studies.
Locations where microgrids can be built are limited to the critical load locations. In Fig. \ref{fig:bn_setup}, line segments that can be hardened via undergrounding are highlighted in green. These portions correspond to segments on three-phase trunks or segments that could form part of an alternate path to supply critical loads. Candidate new lines are indicated with violet dashed lines. Only a small number of these are present, and they are placed only if they are well suited to provide alternate paths of power between a substation and critical loads or between critical loads.

\begin{table}[t!]
\centering
\caption{Problem Setup Properties} \label{tab:psprops}
\begin{tabular}{l r r r}
\hline \hline
Parameter & Case Study 1 & Case Study 2 & Case Study 1 + 2 \\
\hline
Microgrids & 3 & 6 & 9 \\
Candidate Lines & 6 & 6 & 12 \\
Upgradeable Lines & 148 & 462 & 610 \\
\hline \hline
\end{tabular}
\end{table}

The criteria and costs used in the case studies are summarized in Table \ref{tab:global_params}. The critical load--served requirements for each case study are as follows:

\begin{enumerate}
    \item Case Study 1: 98\% 
    \item Case Study 2: 90\% 
    \item Case Study 1 + 2: 90\% 
\end{enumerate}

The amount of noncritical load that must be served varies between 10 and 50\%. Depending on the amount of noncritical load, the optimizer will select a solution that can also provide power to noncritical loads (low amount of noncritical load required to be served) or a solution that adds additional hardening specifically to provide power to a sufficient amount of noncritical loads (high amount of noncritical load required to be served). 

Each potential microgrid location has the choice of 10 possible microgrids sized between 100 KW and 1 MW in 100 KW increments.
As described earlier, the cost of adding microgrids consists of a fixed portion and a capacity-based portion that is linear in the power rating. For the purposes of this study, we perform a sensitivity analysis on the capacity-based portion of the cost.
In practice, this cost (or benefit) could be calculated based on net-present-value (NPV) over the performance period of the microgrid.
For example, given a microgrid power rating of $\sum_{k \in \phases_l} \availableCapacity$ kW, yearly revenue in dollars per kVA $\DGYearlyRevenue$, depreciation rate $\DGDepreciationRate$, and a performance period of $\DGPerformancePeriod$ years, the capacity cost of a microgrid at node $i$ could be calculated as an NPV:

\begin{equation}
    \microgridCost_{l,k} = \sum_{n=0}^\DGPerformancePeriod\left(\frac{1}{1 + \DGDepreciationRate}\right)^n\DGYearlyRevenue \availableCapacity.
\end{equation}

The cost of new lines and upgrading lines remains constant across all design cases studied. Lastly, each of 50 damage scenarios is constructed by damaging lines with a 20\% probability. Note that on average 20\% of lines will be damaged but that the actual number will vary across damage scenarios. 

For Case Study 1 and Case Study 2, we perform a sensitivity analysis on the noncritical load served and microgrid capacity cost parameters. For Case Study 1, 5 values are used for each parameter, giving a total of 25 design cases. For Case Study 2, 9 values are used for each parameter, giving a total of 81 design cases. For the combined Case Study 1 + 2 feeder, a single design case is discussed.

\begin{table}[t!]
\centering
\caption{Constant Parameters for All Case Studies} \label{tab:global_params}
\begin{tabular}{l r}
\hline \hline
Parameter & Value  \\
\hline
Phase Variation & 15\% \\
Critical Load Served & 90\%, 98\% \\
Noncritical Load Served & 10\%, 20\%, ... 50\% \\
Microgrid Sizing Increment\footnotemark[1] & 100 kW \\
Microgrid Fixed Cost & \$25k \\
Maximum Microgrid Size & 1 MW \\
Microgrid Variable Cost\footnotemark[2] & \$100/kW, \$200/kW, ... \$500/kW \\
Line Hardening Cost & \$1M/mile \\
New (Underground) Cost & \$1M/mile + \$25k \\
Damage Probability per Line & 20\% \\
Number of Damage Scenarios & 50 \\
\hline \hline
\end{tabular}

\end{table}

\footnotetext[1]{Power is given per phase.}
\footnotetext[2]{Costs are given in terms of dollars per average power per phase.}

\section{Results}

\begin{figure*}[!t]
\centering
\includegraphics[width=0.9\textwidth]{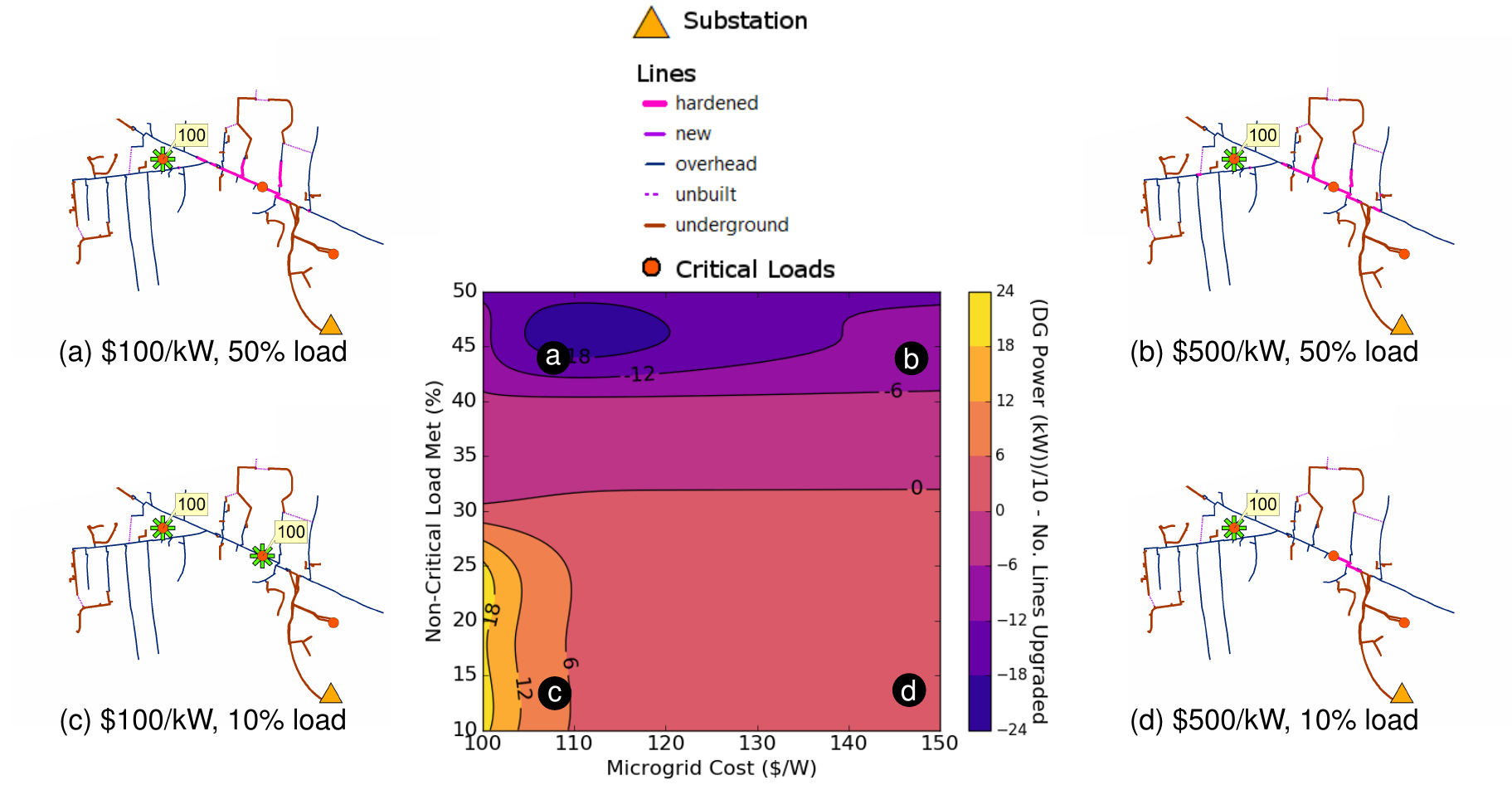}
\caption{Case Study 1 microgrid construction metric and solutions.}
\label{fig:n_contour_soln}
\end{figure*}

Results from the sensitivity analysis for Case Study 1 are shown in Fig. \ref{fig:n_contour_soln}. The center contour plot depicts the difference between the amount of microgrid generation installed in tens of kW and the total number of hardened/new lines. As the amount of noncritical load required to be served increases, the optimizer chooses to harden the feeder trunk. As the cost of installing a microgrid increases, the optimizer chooses to add hardened paths to critical loads instead of installing microgrids. The overall trend from the bottom left to the top right increasingly favors lines over microgrids. The critical load closest to the substation never receives a microgrid because it is connected via a three-phase underground line to the substation and is never damaged under the studied threat scenario. In case (b) at the top right of the figure, one critical load receives neither a microgrid nor a hardened path. The result is reasonable given that this critical load comprises a relatively small portion of the total critical load, i.e., the 98\% critical load served criteria is not violated.

\begin{figure*}[!t]
\centering
\includegraphics[width=0.9\textwidth]{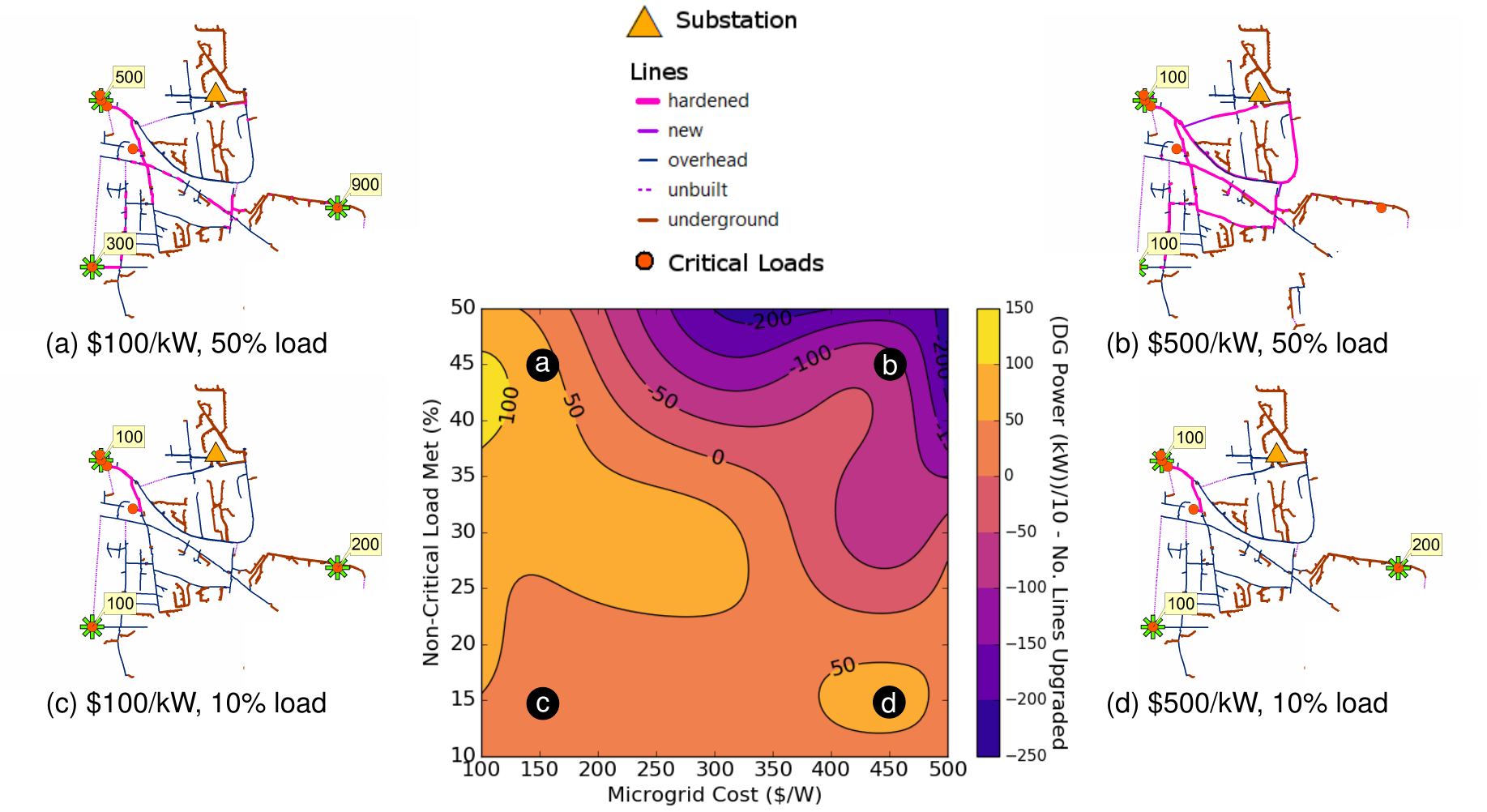}
\caption{Case Study 2 microgrid construction metric and solutions.}
\label{fig:b_contour_soln}
\end{figure*}

Results from the sensitivity analysis for Case Study 2 are displayed in Fig. \ref{fig:b_contour_soln}. The behavior in this  case is a little more complicated. The number of new and hardened lines increases as the noncritical load served increases and the microgrid variable cost increases. However, total microgrid power rating becomes more sensitive with respect to microgrid variable cost as the amount of noncritical load served increases. In case (a), inexpensive microgrid generation is used to provide resilient power to noncritical loads instead of installing new or hardened lines, as shown in case (b). In all cases, the critical loads in the top left of the network figure are connected via hardened lines as a collection of networked microgrids. In case (b), the optimizer chooses to build new lines to provide resilient power to noncritical loads and the critical load at the bottom right of the network diagram. 

\begin{figure}[!t]
\centering
\includegraphics[width=0.45\textwidth]{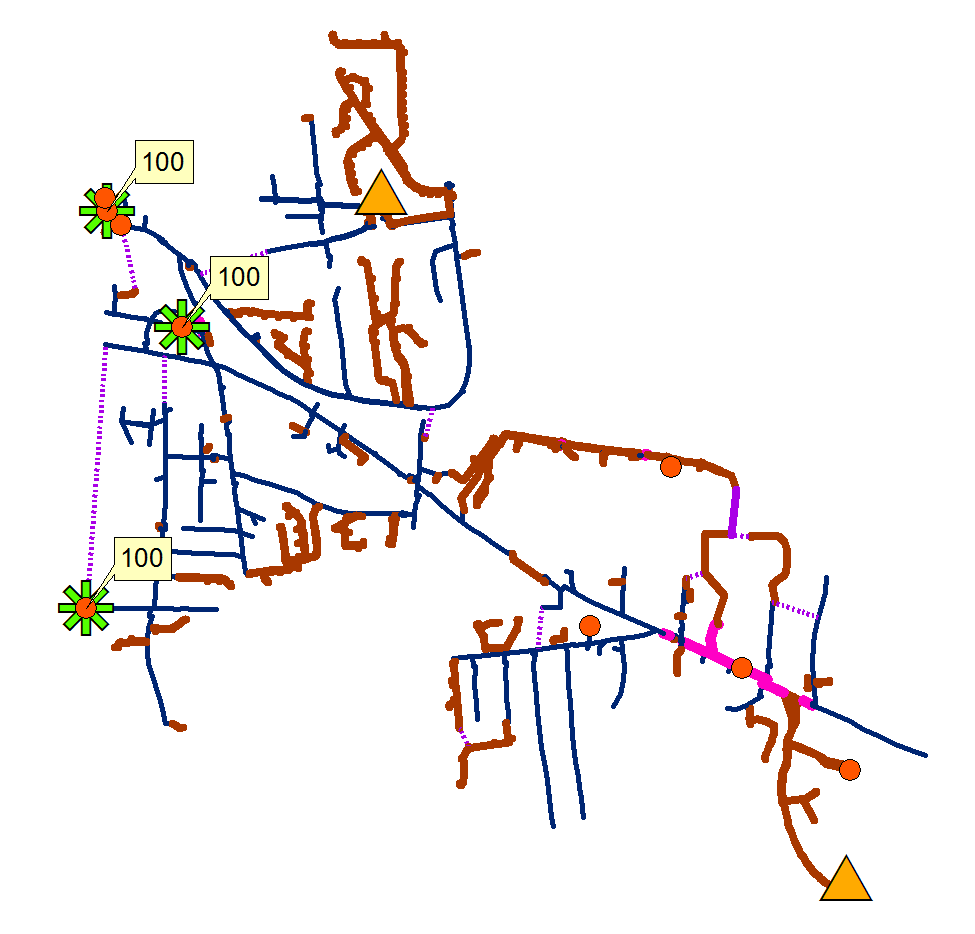}
\caption{Case Study 2 + Case Study 1 solution.}
\label{fig:bn_soln}
\end{figure}

Results for the Case Study 1 + 2 meshed system are illustrated in Fig. \ref{fig:bn_soln} using the following parameters:
\begin{enumerate}
    \item Microgrid variable cost: \$500/kW
    \item Noncritical load served: 33\%
\end{enumerate}

This solution has a smaller set of networked microgrids shown at the top left of the figure. Also of interest is that a new line is constructed to increase meshing of the two feeders, allowing a critical load in Case Study 2 to be supplied resilient power from the Case Study 1 substation with only a small amount of hardening of the Case Study 1 feeder trunk.


The three combinations of feeders are compared in Tables \ref{tab:timecomp} and \ref{tab:costcomp}. For this comparison, the previous parameters of \$500/kW microgrid cost and 33\% noncritical load served are used. Additionally, 90\% of critical load is served for all combinations. This table supports one of the key contributions of this manuscript: \emph{A demonstration on utility data that networked microgrids can bring strong resilience benefits}.  The cost of making each microgrid resilient independently is roughly \$2M. However, meshing the feeders and allowing them to support each other results in a resilient system that costs roughly one third of the independent solutions.

Results from OpenDSS validating that voltage limits are met on the designed solution are presented in Table \ref{tab:voltvalid}.

\begin{table}[t!]
\centering
\caption{Comparison of Calculation Times for Feeders} \label{tab:timecomp} 
\begin{tabular}{l r r r r r}
\hline \hline
Item & Case Study 1 & Case Study 2 & Case Study 1 + 2 \\
 \hline
Avg. Time & 2.91 min & 1.12 h & 2.64 h \\
Std. Dev. & 1.97 min & 1.09 h & \textemdash \\
Worst-Case Time & 6.32 min & 3.25 h & \textemdash \\
No. of Replicates & 81 & 25 & 1 \\
\hline \hline
\end{tabular}
\end{table}

\begin{table}[t!]
\centering
\caption{Comparison of Costs for Feeders} \label{tab:costcomp} 
\begin{tabular}{l r r r r r}
\hline \hline
 Item & Case Study 1 & Case Study 2 & Case Study 1 + 2 \\
 \hline
Cost & \$224k & \$1.72M & \$720k\\
New Lines & 0 & 2 & 1 \\
Hardened Lines & 12 & 82 & 24 \\
No. of Gens & 1 & 1 & 3 \\
Gen Power (kW) & 100 & 100 & 300 \\
\hline \hline
\end{tabular}
\end{table}

\begin{table}[t!]
\centering
\caption{Validation of Voltage Constraints with Unbalanced Distribution System Solver} \label{tab:voltvalid} 
\begin{tabular}{l r r}
\hline \hline
 Case Study &  Lowest Voltage (pu) & Highest Voltage (pu) \\
 \hline
Case Study 1 & 1.019 & 1.030 \\
Case Study 2 & 0.993 & 1.031 \\
Case Study 1 + 2 & 1.026 & 1.030 \\
\hline \hline
\end{tabular}
\end{table}

%% file: Sections/Conclusions.tex
\section{Conclusions}

In this paper we developed a model for assessing the value of introducing networked microgrids to improve distribution system resiliency. We applied 
the approach on two distribution feeders from a northeastern United States electric utility. Under a selection of upgrade options, approximations of power flow physics, an ice/wind fragility model and heuristic search, our approach was demonstrated to scale to problems of practical size.



The results of this study indicate that there is good motivation for networking microgrids for resiliency needs. In particular, the value of networked microgrids is apparent when there are clusters of critical loads that are distant from a substation and the cost of hardening lines is relatively higher than the cost of installing microgrids. It is important to note that when hardening is less expensive, this may reduce the benefit of using networked microgrids. More importantly, perhaps, we also showed the cost of upgrading the resilience of two or more feeders together can be significantly less costly than upgrading the feeder resilience individually. 


There remain a number of interesting future directions. For example, resilient distribution system design remains a computationally difficult problem. Thus, there is room to develop algorithms that improve the performance and scalability of the approach that was discussed here. For example, future work could aggregate lines into upgrade sets where all lines in the set share a single decision variable to harden or build. Similarly, we could allow nonradial operation during extreme events which would cut down on the cycle constraint enumeration. There are also a number of natural parallel computing approaches that could exploit the decomposable structure of this problem.  Finally, communication and control are increasingly becoming a part of distribution system design and these will play a critical role in the future resilience of distribution feeders. Thus, there is a need for model development that includes communication pathways and controls.




\section*{Acknowledgements}
The work was funded by the DOE-OE Smart Grid R\&D Program
in the Office of Electricity in the US Department of
Energy. 